\documentclass[11pt, letterpaper]{article}

\usepackage[left]{lineno}

\usepackage[T1]{fontenc}
\usepackage{lmodern}
\usepackage{amsmath, amssymb, amsthm, dsfont, mathtools}
\usepackage{thmtools}
\usepackage{thm-restate}
\usepackage{xspace}
\usepackage{graphicx}
\usepackage{tikz}
\usepackage{enumitem}
\usetikzlibrary{decorations.pathmorphing}
\usetikzlibrary{arrows.meta,positioning,calc}
\usetikzlibrary{decorations.pathreplacing}
\usepackage[dvipsnames]{xcolor}
\definecolor{darkgreen}{RGB}{0,128,43}
\usepackage[ruled,vlined, linesnumbered]{algorithm2e}
\usepackage{enumitem}

\title{Sublogarithmic Online Graph Exploration}
\author{Júlia Baligács\thanks{Email: \href{mailto:jbaligacs@gmail.com}{jbaligacs@gmail.com}. Funded by the European Union (ERC, CCOO, 101165139).
Views and opinions expressed are however those of the author only and do not necessarily reflect those of the European Union or the European Research Council. Neither the European Union nor the granting authority can be held responsible for them.}
\vspace{2mm}\\ University of Oxford}
\date{}

\usepackage[margin=1in]{geometry}

\usepackage{todonotes}
\usepackage{hyperref}
\usepackage[nameinlink]{cleveref}

\crefformat{lemma}{#2Lemma~#1#3}
\crefformat{theorem}{#2Theorem~#1#3}
\crefformat{proposition}{#2Proposition~#1#3}
\crefformat{remark}{#2Remark~#1#3}
\crefformat{equation}{#2(#1)#3}
\crefformat{corollary}{#2Corollary~#1#3}
\crefformat{example}{#2Example~#1#3}
\crefformat{definition}{#2Definition~#1#3}
\crefformat{figure}{#2Figure~#1#3}
\crefformat{observation}{#2Observation~#1#3}
\crefformat{section}{#2Section~#1#3}
\crefformat{question}{#2Question~#1#3}
\crefformat{algorithm}{#2Algorithm~#1#3}

\definecolor{LinkViolet}{hsb}{0.37,1,0.5}
\hypersetup{
  colorlinks=true,
  linkcolor=Violet,
  citecolor=Violet,
  urlcolor=green!40!black
}

\newcommand{\Z}{\mathbb{Z}}

\newcommand{\R}{\mathbb{R}}
\newcommand{\bigO}{\mathcal{O}}
\renewcommand{\subset}{\subseteq}
\newcommand{\opt}{\textsc{Opt}}
\newcommand{\alg}{\textsc{Alg}}

\theoremstyle{definition}
\newtheorem{definition}{Definition}
\newtheorem{remark}[definition]{Remark}

\theoremstyle{plain}
\newtheorem{theorem}[definition]{Theorem}

\newtheorem{lemma}[definition]{Lemma}

\newtheorem{observation}[definition]{Observation}

\newtheorem{question}[definition]{Question}

\newcommand{\Explore}{\ensuremath{\textsc{Threshold-hDFS}}\xspace}
\newcommand{\component}{\mathrm{comp}}

\usepackage{tikz}
\usepackage[ruled,vlined, linesnumbered]{algorithm2e}
\usetikzlibrary{arrows.meta,decorations.markings}
\usetikzlibrary{decorations.pathreplacing}

\begin{document}

\maketitle

\begin{abstract}
We study the online graph exploration problem, in which an agent must visit every vertex of an initially unknown weighted graph on $n$ vertices and return to its starting vertex.
In this model, a vertex's incident edges and their weights are revealed when it is first visited.
Despite extensive research, the best known competitive guarantee for general graphs had remained~$\mathcal{O}(\log n)$, a bound already established in 1977 for the greedy TSP heuristic Nearest Neighbor.
Subsequent approaches, including hierarchical depth-first search (hDFS) and Blocking, achieve better guarantees on restricted graph classes but have logarithmic or worse competitive ratios on general graphs.
This raises the question of whether the logarithmic barrier is inherent and Nearest Neighbor is asymptotically optimal.

We answer this question in the negative by giving an $\mathcal{O}(\log n/\log \log n)$-competitive algorithm.
Our algorithm is a slight modification of hDFS, and we establish its competitive guarantee with a surprisingly short proof.
\end{abstract}

\section{Introduction}

Online graph exploration models various real-world tasks that involve navigating and exploring unknown environments.
Examples include a robot mapping unfamiliar terrain, a rescue worker searching an unfamiliar area, or a web crawler discovering webpages for indexing.
In such settings, the environment is revealed only as exploration proceeds, so navigation decisions must be made using incomplete information.
More generally, finding efficient routes is of fundamental theoretical and practical importance and has motivated decades of extensive research on approximation algorithms for the traveling salesperson problem~(TSP)~\cite{Arora98,BartalGK12,Christofides76,GurvitsKL24,KarlinKO21,Klein05,MomkeS11,OveisGharanSS11,SvenssonTV20,TraubV20}.
In online graph exploration, we study approximation algorithms for TSP with limited information.

Formally, in the online graph exploration problem, introduced by Kalyanasundaram and Pruhs in~1994~\cite{Kalyanasundaram94}, a single agent needs to explore a connected, undirected graph~$G=(V,E,w)$ with non-negative edge weights $w\colon E\to\R_{\geq 0}$.
The agent is initially located at a \emph{starting position} $s\in V$.
The graph is initially unknown: upon visiting a vertex, the agent learns the unique identifiers of its neighbors and the weights of the corresponding edges.
At any time during exploration, we call a previously visited vertex \emph{explored} and an unvisited vertex adjacent to an explored vertex a \emph{boundary vertex} (i.e., it is already revealed but not yet visited).
A \emph{boundary path} is a path $(v_1,\dots,v_k)$ such that $v_1,\dots,v_{k-1}$ are explored and $v_k$ is a boundary vertex.
At each step, the agent chooses a boundary vertex and traverses a boundary path from its current position to that vertex, incurring a cost equal to the path's length.
After visiting all vertices, the agent must return to~$s$.
The objective is to minimize the total traversal cost.

For a deterministic online graph exploration algorithm $\alg$, let $\alg(G,s)$ denote its total traversal cost on graph $G$ with starting vertex $s\in V(G)$.
We denote by $\opt(G)$ the \emph{offline optimal cost}, that is, the length of a shortest TSP tour.
Note that this value is independent of the starting position.
We say that $\alg$ is \emph{$\rho$-competitive} if $\alg(G,s)\leq\rho\,\opt(G)$ for every graph~$G$ and starting position $s\in V(G)$.\footnote{This is the \emph{strict} notion of competitiveness.
The non-strict notion allows $\alg(G,s)\leq\rho\,\opt(G)+C$ for a constant $C$ independent of $G$ and $s$.
Both notions are known to yield the same competitive ratio in online graph exploration~\cite{Baligacs26}, so we omit the qualifier ``strict''.}
The \emph{competitive ratio} of $\alg$ is the infimum over all $\rho$ for which $\alg$ is $\rho$-competitive.

The central question in online graph exploration is the following.

\begin{question}[Kalyanasundaram and Pruhs, 1994~\cite{Kalyanasundaram94}]
Is there a constant-competitive algorithm for the online graph exploration problem?
\end{question}

In other words, this asks whether TSP can be approximated within a constant factor when the graph is revealed only during traversal.
Despite extensive research~\cite{BaligacsDHS26,BaligacsHV26,Baligacs26,BirxDHK21,BockenhauerFU22,TadpoleGraphs,DobrevKM12,EberleLMNS22,Fritsch21,Gehnen26,HougardyW15,Kalyanasundaram94,Unicyclic,Megow2012,miyazaki}, this question remains open to this day.
Prior to our work, the best known upper bound on general graphs was $\bigO(\log n)$, where $n$ denotes the number of vertices.
The currently best known lower bound is~4~\cite{Baligacs26}.

The $\bigO(\log n)$ bound is achieved by the \emph{Nearest Neighbor} algorithm, which repeatedly traverses a shortest (boundary) path to a closest boundary vertex.
This greedy strategy was already analyzed as a simple TSP heuristic in~1977, way before the online graph exploration problem was introduced, and its competitive ratio is $\Theta(\log n)$~\cite{Rosenkrantz77}.
The logarithmic lower bound was later strengthened to already hold on unweighted planar graphs~\cite{HougardyW15} and on weighted trees~\cite{Fritsch21}.
Alternative strategies include \emph{hierarchical depth-first search} (hDFS) and \emph{Blocking}, introduced by Megow, Mehlhorn, and Schweitzer~\cite{Megow2012}, where the latter algorithm is an adaptation of the ShortCut algorithm introduced by Kalyanasundaram and Pruhs~\cite{Kalyanasundaram94}.
These algorithms perform better than Nearest Neighbor on some restricted graph classes, but not on general graphs:
hDFS is constant-competitive on graphs with a bounded number of distinct edge weights~\cite{Megow2012}.
Blocking achieves a constant competitive ratio on bounded-genus graphs~\cite{Megow2012}, and was later shown to be constant-competitive on every graph class excluding a fixed minor~\cite{BaligacsDHS26}.
Further work establishes optimal competitive ratios for cycles and unweighted graphs~\cite{miyazaki} and for tadpole graphs~\cite{TadpoleGraphs}, as well as constant competitive upper bounds for unicyclic graphs~\cite{Fritsch21,Unicyclic} and cactus graphs~\cite{Fritsch21}.
On general graphs, however, the standard rounded version of hDFS has competitive ratio $\Theta(\log n)$~\cite{Megow2012}, while Blocking with a fixed `blocking parameter' $\delta$ has a lower bound of $\Omega(n^{1/(\delta+4)})$ on its competitive ratio~\cite{Megow2012}.
Allowing the blocking parameter to depend on $n$, a lower bound of $\Omega(\log n)$ was established for a broad range of parameter choices~\cite{BaligacsDHS26}.

To summarize, despite decades of research and substantially different algorithmic approaches and progress on restricted graph classes, the logarithmic guarantee of Nearest Neighbor had resisted improvement on general graphs, suggesting that $\Theta(\log n)$ might be an inherent barrier.
The greedy principle underlying Nearest Neighbor is particularly compelling on general graphs: Intuitively, without any additional structural assumptions, it is difficult (and unclear whether it is even possible) to anticipate when a more expensive immediate decision could reduce the overall exploration cost.
This raises a more basic question: can a competitive ratio of $o(\log n)$ be achieved on general graphs, or is Nearest Neighbor asymptotically optimal?

In this work, we resolve this question by giving an algorithm with a competitive ratio of~$o(\log n)$.
More precisely, we prove the following.

\begin{theorem}
\label{thm:main}
There exists an algorithm for online graph exploration with a competitive ratio of~$\bigO(\log n/\log \log n)$.
\end{theorem}

This gives the first asymptotic improvement on the general-graph competitive guarantee since 1977 and establishes that Nearest Neighbor is \emph{not} asymptotically optimal.

Our algorithm is a slight modification of hDFS, and the proof of its competitive guarantee is surprisingly short.

\paragraph*{Further related work.}

As briefly mentioned above, research on the online graph exploration problem has mostly focused on competitive guarantees on restricted graph classes.
As a result, it is known that the competitive ratio is 2 on unweighted graphs~\cite{miyazaki}, $(1+\sqrt{3})/2\thickapprox 1.366$ on cycles~\cite{miyazaki},  at most~2.5 on unicyclic graphs~\cite{Fritsch21,Unicyclic}, at most~$5/2+\sqrt{2}\approx3.914$ on cactus graphs~\cite{Fritsch21}, at most $2k$ on graphs with $k$ distinct edge weights~\cite{Megow2012}, at most 16 on planar graphs~\cite{Kalyanasundaram94, Megow2012}, and constant on any graph class excluding a fixed minor~\cite{BaligacsDHS26} (which includes graph classes of bounded genus or bounded treewidth).
A recent work considers randomized algorithms for the problem and establishes an improved upper bound on the expected competitive ratio of~$\thickapprox 1.314$ on cycles~\cite{BaligacsHV26}.
The lower bound for online graph exploration was sequentially improved from $5/2$~\cite{DobrevKM12} to $10/3$~\cite{BirxDHK21}, with the currently best lower bound being~4~\cite{Baligacs26} up to this day. All of these lower bound constructions already hold on the class of planar graphs.

Another line of work studies how additional information can improve exploration.
In the advice model, an oracle with full knowledge of the graph provides information to the algorithm.
Dobrev, Kr{\'a}lovi{\v{c}}, and Markou~\cite{DobrevKM12} show that $\bigO(n)$ bits of advice suffice to achieve constant competitiveness, while B\"ockenhauer, Fuchs, and Unger~\cite{BockenhauerFU22} give an algorithm achieving optimal exploration with $\bigO(n+m)$ bits of advice.
Eberle et al.~\cite{EberleLMNS22} study learning-augmented exploration with predicted tours or spanning trees, obtaining constant competitive guarantees for accurate predictions while retaining an $\bigO(\log n)$ guarantee for arbitrary predictions.
Gehnen, Klasing, and Naquin~\cite{Gehnen26} consider a model in which the graph topology is known in advance but edge weights are initially available only through estimates, and analyze competitiveness in terms of the uncertainty in these estimates.

There are several related exploration settings studied in the literature.
Exploration of directed graphs has been studied both when every edge must be traversed and when the objective is to visit every vertex.
For strongly connected directed graphs, Foerster and Wattenhofer~\cite{FoersterW16} establish sharp competitive ratios of $n-1$ for weighted vertex exploration and $\frac{n}{2}+\frac12-\frac1n$ for unweighted vertex exploration.
In the edge-exploration variant on unweighted graphs, where edge destinations are revealed only upon traversal, the best known upper bound is $\bigO(d^8)$~\cite{FleischerT05} and the best known lower bound is $\Omega(d)$~\cite{DengPapadimitriou1999}, where the Eulerian deficiency $d$ is the minimum number of edges that must be added to make the graph Eulerian.
In \emph{collaborative exploration}, a team of $k$ agents, each moving at unit speed, need to explore the graph, with the objective of minimizing the time until all vertices are visited by some agent.
Here, it is already unclear how to optimally explore unweighted trees.
After a long line of research~\cite{Cosson24,CossonMas24,CossonMV23,dereniowski,FraigniaudGKP06,OrtolfS14}, Cosson and Massoulié establish in a recent work an algorithm with a competitive ratio of $\bigO(\log^2 k)$~\cite{CossonMas26}, while the best known lower bound is~$\Omega(\log k/ \log \log k)$~\cite{disserlower}.

The offline counterpart to online graph exploration is metric TSP.
The problem is NP-hard, but its importance for applications in logistics and planning, together with immense theoretical interest, has motivated extensive research on approximation algorithms for metric TSP over several decades.
Christofides' classical $3/2$-approximation~\cite{Christofides76}, introduced in 1976, remained the best known guarantee for more than four decades.
Karlin, Klein, and Oveis Gharan~\cite{KarlinKO21} improved this factor to $3/2-\varepsilon$ for some small constant $\varepsilon>0$, with a subsequent refinement in~\cite{GurvitsKL24}.
Additional structure in the metric permits substantially stronger guarantees: polynomial-time approximation schemes are known for Euclidean metrics~\cite{Arora98} and, more generally, for metrics of bounded doubling dimension~\cite{BartalGK12}.
Graph structure can likewise be exploited: shortest-path metrics of weighted planar graphs admit a polynomial-time approximation scheme~\cite{Klein05}, while graphic TSP, defined by shortest-path distances in unweighted graphs, admits better approximation factors than those known for general metrics~\cite{MomkeS11,OveisGharanSS11}.
Allowing the cost of traversing an edge to depend on the direction yields asymmetric TSP, for which constant-factor approximations are known~\cite{SvenssonTV20,TraubV20}.

\paragraph*{Outline.}
In \cref{sec:algorithm}, we introduce the online graph exploration algorithm $\Explore_b$, parameterized by a value $b\in \R_{> 1}$, and observe some of its key properties.
The algorithm is a slight adaptation of the algorithm hDFS introduced in~\cite{Megow2012}, and, at the end of \cref{sec:algorithm}, we explain the differences between the two algorithms, which allow us to obtain a better competitive ratio than established in previous work.

In \cref{sec:competitive_analysis}, we analyze the competitive ratio of $\Explore_b$, depending on the parameter~$b$.
In that section, we will also see that our analysis is tight and, more importantly, that the choice $b=\sqrt{\log n}$ yields the desired competitive ratio of $\bigO(\log n/ \log \log n)$.
The issue, however, is that the agent does not know the number of vertices $n$ in advance, so it is unclear how to set the parameter~$b$.
In \cref{sec:doubling}, we show how to circumvent this issue by employing a doubling strategy, so that $b$ is gradually increased as more vertices are explored.
This then completes the proof of~\cref{thm:main}.

\section{The algorithm}
\label{sec:algorithm}

In this section, we introduce the algorithm $\Explore_b$.
For this, we begin with collecting the notation that we need.

\subsection{Preliminaries and notation}

For a graph $G=(V,E)$ with edge weights $w \colon E \to \R_{\geq 0}$, define $G_W:=(V,\{e\in E: w(e)\leq W\})$ to be the graph restricted to edges of weight at most $W$.
Define $\component_G(u,W)$ as the connected component of $G_W$ that contains~$u$, which we identify with the set of vertices in it.
Let us make two basic but crucial observations.
First, at any time during exploration, if the agent is located at position $u$, it knows whether $\component_G(u,W)$ is already explored.
Indeed, $\component_G(u,W)$ is not yet fully explored if and only if there exists a boundary path from $u$ such that all of its edges have weight at most $W$.
Second, for any $W'<W$, a connected component $C$ of $G_W$ partitions into connected components of $G_{W'}$ (i.e., every component of~$G_{W'}$ is either fully contained in $C$ or disjoint from $C$), resulting in a hierarchical structure.
We will sometimes consider the \emph{quotient graph} of $C$ induced by the partition of $G_{W'}$.
This is the graph that has a vertex for every connected component of $G_{W'}$ in $C$, and two vertices are connected if the corresponding components are connected by an edge of weight at most~$W$.
Note that such an edge has a weight in $(W',W]$.

In the algorithm, we will sometimes compute a TSP tour (i.e., a closed walk visiting all vertices) of an already fully explored component $C=\component_G(u,W)$.
We write a TSP tour of a component on $k-1$ vertices as a permutation of the vertices with the starting vertex appended to the end, i.e., $(x_1,\dots, x_k)$ with $x_1=x_k$ and $\{x_1,\dots, x_{k-1}\}=C$.
In particular, when the component consists of a single vertex $v$, we write the TSP tour of the component as~$(v,v)$.
The total length of the tour is then $\sum_{i=1}^{k-1} d(x_i,x_{i+1})$.

In our algorithm, a constant-factor approximation to a shortest TSP tour will suffice.
Note that, if $T$ is a minimum spanning tree (MST) of a graph $G$, we have
\begin{equation*}
w(T)\leq \opt (G)\leq 2w(T),
\end{equation*}
since doubling the edges of an MST gives a factor-2 approximation to a shortest TSP tour.
In the analysis, we therefore bound the exploration cost in terms of $w(T)$, allowing us to make use of structural properties of an~MST.

Last, a \emph{boundary edge} is an edge with one explored and one unexplored endpoint (i.e., a boundary path consisting of one edge).
Moreover, given a set of vertices $S\subset V(G)$, by $G[S]$, we denote the induced subgraph of $G$ on vertex set $S$.

\subsection{Algorithm definition}

\begin{algorithm}[t]
\caption{$\Explore_b(u,W)$}
\label{alg:explore}
\small

\KwIn{agent's current position $u$; threshold $W\in b^{\mathbb Z}\cup\{0,\infty\}$}

\If{$W=0$}{
    explore $\component_G(u,0)$ by DFS using only zero-weight edges, returning to $u$\;
    \Return\;
}

$W'\gets\min\{Y\in b^{\mathbb Z}\cup\{0\}:\component_G(u,Y)\text{ is not fully explored}\}$

\While{$W'< W$}{
    $\Explore(u,W')$\;
    Recompute $W'$\;
}
compute a TSP tour $T=(x_1,\dots,x_k)$ of $G[\component_G(u,W/b)]$ of weight at most~$2w(\mathrm{MST}(G[\component_G(u,W/b)]))$ with $x_1=x_k=u$

\For{$i=1,\dots, k-1$}{
    \While{there is a boundary edge $e=\{x_i,y\}$ with $w(e)\leq W$}{
        traverse $e$\\
        $\Explore(y,W)$\\
        traverse $e$ back to $x_i$\\
    }
	traverse a shortest path in $G[\component_G(u,W/b)]$ from $x_i$ to $x_{i+1}$
}

\end{algorithm}

Now, we have all the preliminaries in place to define the algorithm $\Explore$, which is formally defined in \cref{alg:explore}.
The algorithm is a slight modification of hDFS introduced in~\cite{Megow2012}.
We explain the differences at the end of this section.

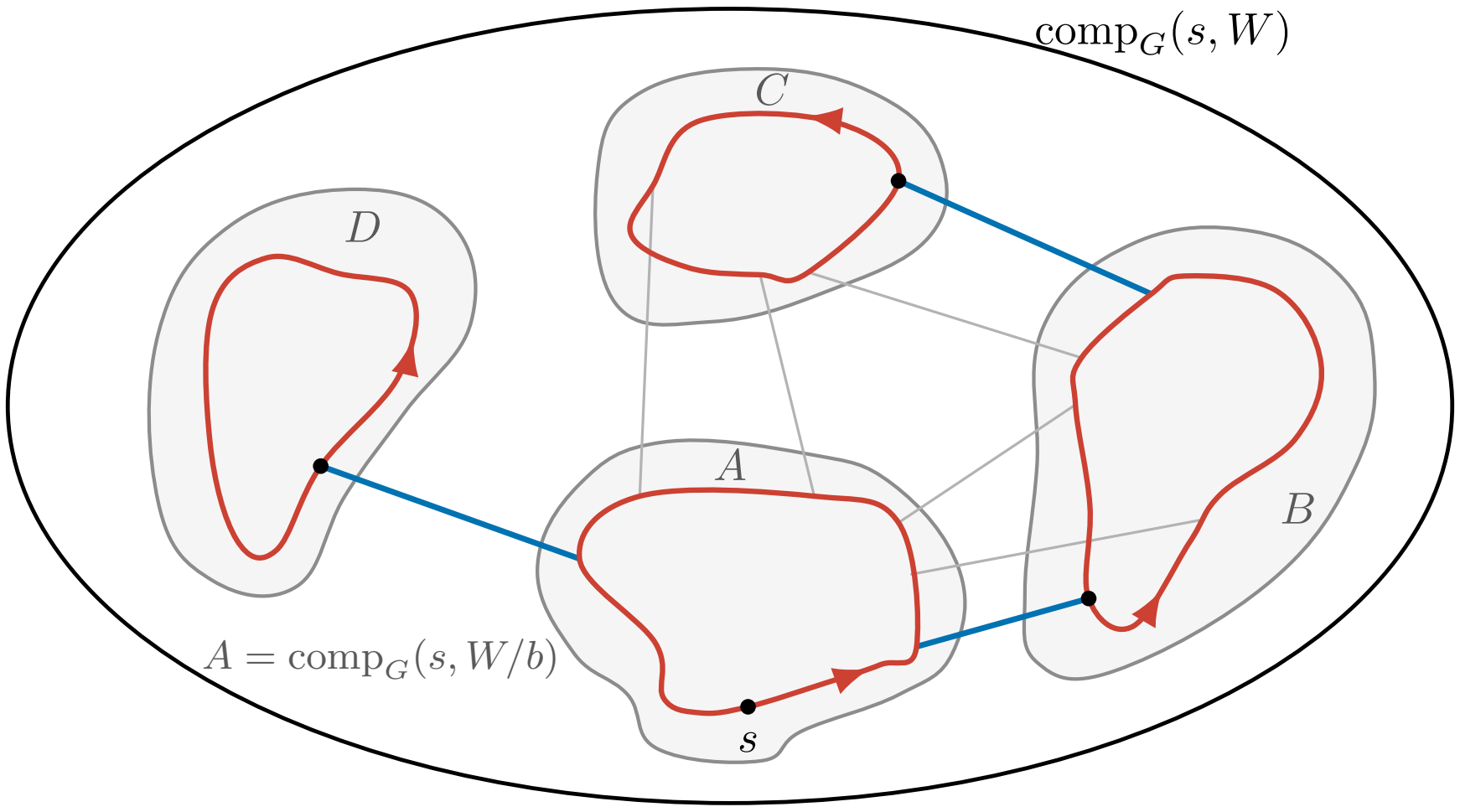
\begin{figure}[t]
\centering
\begin{tikzpicture}[scale=0.95, font=\small]
\definecolor{tourcolour}{RGB}{204,65,50}
\definecolor{dfscolour}{RGB}{0,114,178}
\tikzset{
  component/.style={draw=black!45,fill=black!4,line width=.8pt},
  adjacency/.style={draw=black!30,line width=.6pt},
  extension/.style={draw=dfscolour,line width=1.3pt},
  tour/.style={
    draw=tourcolour,line width=1.2pt,
    postaction={decorate},
    decoration={
  markings,
  mark=at position .05 with {\arrow{Latex}}
}
  },
  vertex/.style={circle,fill=black,inner sep=0pt,minimum size=3.5pt}
}

\draw[line width=.9pt] (0,.15) ellipse[x radius=6cm,y radius=3.3cm];
\node[fill=none,inner sep=2pt] at (3.6,3.25) {$\mathrm{comp}_G(s,W)$};

\draw[component] plot[smooth cycle,tension=.7] coordinates {
  (-1.6,-1.2) (-1.25,-.5) (-.55,-.15) (.65,-.25)
  (1.35,-.5) (1.9,-1.2) (1.85,-1.9) (1.4,-2.25)
  (.6,-2.55) (.25,-2.8) (-.6,-2.7) (-.85,-2.25) (-1.3,-1.9)
};
\draw[component] plot[smooth cycle,tension=.7] coordinates {
  (-4.75,-.65) (-4.75,.75) (-4.15,1.65) (-3.1,1.95)
  (-2.3,1.65) (-2.15,.85) (-2.7,.1) (-3.2,-.6)
  (-3.6,-1.35) (-4.3,-1.3)
};
\draw[component] plot[smooth cycle,tension=.7] coordinates {
  (-.95,1) (-1.1,2.1) (-.75,2.75) (.3,2.95)
  (1.25,2.72) (1.75,2.2) (1.65,1.55) (.8,1.1) (-.15,.85)
};
\draw[component] plot[smooth cycle,tension=.7] coordinates {
  (2.45,-1.4) (2.55,-.35) (2.6,.8) (3.35,1.5)
  (4.35,1.6) (5.1,1.2) (5.35,.5) (5.2,-.35)
  (4.5,-1.3) (3.2,-2.05) (2.55,-1.98)
};

\coordinate (s) at (.15,-2.35);
\coordinate (aD) at (-1.25,-1.12);
\coordinate (dA) at (-3.4,-.35);
\coordinate (aB) at (1.55,-1.85);
\coordinate (bA) at (2.98,-1.45);
\coordinate (bC) at (3.5,1.08);
\coordinate (cB) at (1.4,2.02);

\draw[adjacency] (-.75,-.6) -- (-.64,1.98);
\draw[adjacency] (.7,-.6) -- (.25,1.24);
\draw[adjacency] (1.4,-.82) -- (2.87,.16);
\draw[adjacency] (1.5,-1.25) -- (3.9,-.8);
\draw[adjacency] (.65,1.26) -- (2.92,0.55);

\draw[extension] (aD) -- (dA);
\draw[extension] (aB) -- (bA);
\draw[extension] (bC) -- (cB);

\draw[tour] plot[smooth cycle,tension=.65] coordinates {
  (s) (.8,-2.15) (1.25,-2) (aB) (1.4,-.82)
  (.7,-.6) (-.75,-.6) (aD) (-.62,-1.82) (-.55,-2.28) (-.25,-2.4)
};
\draw[tour] plot[smooth cycle,tension=.65] coordinates {
  (dA) (-2.75,.4) (-2.65,1.08) (-3.25,1.25)
  (-3.85,1.38) (-4.3,.95) (-4.32,-.15) (-4.1,-.96) (-3.77,-1.05)
};
\draw[tour] plot[smooth cycle,tension=.65] coordinates {
  (cB) (.9,2.5) (-.25,2.52) (-.64,1.98)
  (-.82,1.57) (-.35,1.3) (.25,1.24) (.65,1.26)
};
\draw[tour] plot[smooth cycle,tension=.65] coordinates {
  (bA) (3.35,-1.68) (3.82,-.98) (4.08,-.58) (4.7,-.12)
  (4.91,.5) (4.55,1.1) (3.8,1.23) (bC) (2.92,.55) (2.87,.16) (2.99,-.68)
};
\foreach \v in {bA,cB,dA}
  \node[vertex] at (\v) {};
\node[vertex,label={[label distance=0pt]below:$s$}] at (s) {};

\node[text=black!65] at (0,-.34) {$A$};
\node[text=black!65] at (4.72,-.70) {$B$};
\node[text=black!65] at (.35,2.78) {$C$};
\node[text=black!65] at (-3.05,1.64) {$D$};
\node[font=\footnotesize,text=black!65] at (-2.9,-1.95) {$A=\mathrm{comp}_G(s,W/b)$};
\end{tikzpicture}
\caption{Hierarchical exploration at threshold~$W$.
The gray regions are components of $G_{W/b}$ within $\mathrm{comp}_G(s,W)$.
Red curves represent tours within these components (computed in line 8 of~\cref{alg:explore}).
The gray edges represent edges of $G_W$ between components of $G_{W/b}$, i.e., they have weight in $(W/b,W]$.
The blue edges represent edges traversed due to line 11 of~\cref{alg:explore}.
They form a DFS tree of the quotient graph.
The black points represent the first vertex visited in each of the components.}
\label{fig:hierarchical-dfs}
\end{figure}

The algorithm is parameterized by $b\in \R_{> 1}$
and the  high-level intuition is that we explore $\component_G(s,W)$ for increasing values of~$W$, where $W$ ranges over powers of $b$, until the entire graph is explored.
When $\component_G(u,W/b)$ is already explored, we expand the explored part to $\component_G(u,W)$ as follows (cf.~\cref{fig:hierarchical-dfs}): We follow a TSP tour of $\component_G(u,W/b)$ and, when we encounter a boundary edge of weight at most $W$, we traverse it and then recursively invoke the algorithm with threshold $W$.

A bit more formally, the algorithm $\Explore_b$ takes as input a vertex $u$ denoting the agent's current position and a threshold $W$.
At the end of execution of $\Explore_b(u,W)$, the agent will have fully explored $\component_G(u,W)$.
Lines 1-3 (cf.~\cref{alg:explore}) solve the special case where~$W=0$, where we simply explore the component with depth-first search, incurring zero cost.
Lines~4--7 ensure that $\component_G(u,W/b)$ is fully explored before we expand.
For this, we identify the smallest value of the threshold, for which the component is not explored (in line 4), where we employ the convention that $\min \emptyset = \infty$, and increase the threshold until  $\component_G(u,W/b)$ is fully explored.
When reaching line 8, we have therefore fully explored  $\component_G(u,W/b)$ and can compute a TSP tour of it.
Here, a constant-factor approximation to the optimum in $G[\component_G(u,W/b)]$ (e.g., a depth-first search traversal of an MST) suffices for our purposes.
We follow the TSP tour and traverse a boundary edge of weight at most~$W$ when encountering one, which leads us to a new connected component of $G_{W/b}$, where we proceed recursively.
Once the recursive exploration is complete, we backtrack and resume the TSP tour.

Consider \cref{fig:hierarchical-dfs} for an illustrative example. 
Here, $\component_G(s,W)$ consists of four components $A,B,C,D$ of $G_{W/b}$, with $s\in A$.
The agent first explores $A$ at threshold~$W/b$ and then follows its red tour until traversing the blue edge to~$B$.
On reaching $B$, it explores this component at threshold~$W/b$ before following its red tour and interrupting it by following the blue edge to~$C$.
After repeating this procedure in $C$, it backtracks to $B$, completes its tour, and returns to $A$.
It resumes the tour of $A$, explores $D$ in another excursion, and returns to finish the tour at~$s$.
Note that the blue edges form a depth-first search trajectory on the quotient graph, which gives the algorithm its name hierarchical depth-first search.

We note down two properties of the algorithm that are immediate from its definition.

\begin{samepage}
\begin{observation}
\label{obs:algorithm_basics}
Let $b>1$, $G=(V,E,w)$ be a graph, and $s\in V$.
\begin{enumerate}[label=\alph*)]
\item The algorithm $\Explore_b(s,\infty)$ is a correct online graph exploration algorithm, i.e., it terminates, visits all vertices, the agent returns to the starting position, and all steps are well-defined (in particular, whenever line 8 is executed, $\component_G(u,W/b)$ is fully explored).
\item Let  $W\in b^{\mathbb Z}$ and $u\in V$. Consider the first time that $u$ is visited. Then $\component_G(u,W/b)$ will be fully explored before an edge of weight more than $W/b$ is traversed. \label{item:obs_fully_explore}
\end{enumerate}
\end{observation}
\end{samepage}

\paragraph*{Comparison to hDFS.}
Our algorithm \Explore differs from hDFS in the following key respect.
Whereas our thresholds $W$ are powers of $b$, hDFS uses all edge weights as thresholds.
In the rounded version of hDFS analyzed in \cite{Megow2012}, the edge weights are first rounded up to powers of a fixed constant base.
If all edge weights were rounded up to powers of $b$, this distinction between the threshold rules would disappear.
However, it is crucial for our algorithm to use the original edge weights when computing the TSP tours in line~8.
Using rounded weights in this step would introduce an additional factor of $b$ in the bound on the touring cost.
Since we allow $b$ to depend on $n$, we cannot afford this additional factor.

The remaining differences between \Explore and hDFS are cosmetic.


\section{Competitive analysis}
\label{sec:competitive_analysis}

In this section, we prove the following result.

\begin{theorem}
\label{thm:hDFS_b}
The algorithm $\Explore_b$ is $\bigO \bigl(b+\frac{\log(n)}{\log(b)} \bigr)$-competitive on graphs with~$n$ vertices.
\end{theorem}

Observe that, if the number of vertices $n$ is known in advance to the agent, the theorem already implies the existence of an $\bigO(\log n/\log \log n)$-competitive algorithm as desired for \cref{thm:main}.
Indeed, if $n$ is known in advance, we can simply set $b=\sqrt{\log n}$ (for sufficiently large $n$) and obtain a competitive ratio of $\bigO(\sqrt{\log n} + \log n/ (0.5 \log \log n))= \bigO(\log n/\log \log n)$.

In this section, we analyze the algorithm for a fixed parameter $b$, which may be chosen as a function of $n$.
In the next section, we use a simple ``doubling strategy'' that gradually increases~$b$ as exploration proceeds.
This allows us to achieve a competitive ratio of~$\bigO(\log n/\log \log n)$ without knowing $n$ in advance.

For the remainder of this section, let $G=(V,E,w)$ be an arbitrary fixed graph, $s\in V$,  and $T$ be a minimum spanning tree of $G$.
The goal is to bound the total cost incurred by $\Explore_b(s,\infty)$ on graph $G$ in terms of $w(T)$.
We assume throughout that $w(T)=\sum_{e\in T} w(e)>0$ as otherwise, there is nothing to show.

\subsection{Hierarchical component structure and minimum spanning trees}

We begin with noting down the basic graph-theoretic ingredients that we will use in our analysis of the algorithm.
The following lemma is a simple consequence of $T$ being a minimum spanning tree.

\begin{lemma}
\label{lem:components_mst}
For every vertex $u\in V(G)$ and every $W\in \R_{\geq 0}$, we have $\component_T(u,W)=\component_G(u,W)$.
Moreover, $T[\component_G(u,W)]=T_W[\component_G(u,W)]$ is a minimum spanning tree of $G[\component_G(u,W)]$.
\end{lemma}

\begin{proof}
Since $T$ is a subgraph of $G$, we have $\component_T(u,W)\subseteq \component_G(u,W)$.
Assume for a contradiction that $\component_T(u,W)\subsetneq \component_G(u,W)$.
Then $\component_G(u,W)$ contains multiple components of $T_W$, and there is an edge of $G_W$ connecting $\component_T(u,W)$ to another such component.
That is, there exists an edge $e=\{x,y\}\in E(G)$ with $w(e)\leq W$, $x\in\component_T(u,W)$, and $y\in \component_G(u,W)\setminus\component_T(u,W)$.
The unique path in $T$ between $x$ and $y$ contains an edge of weight greater than $W$, since the two vertices lie in distinct components of $T_W$.
Therefore, $T\cup\{e\}$ contains a cycle in which $e$ is strictly lighter than the heaviest edge.
Replacing that edge by $e$ yields a spanning tree of strictly smaller weight, contradicting the minimality of $T$.

It follows that $T[\component_G(u,W)]=T_W[\component_G(u,W)]$ is a spanning tree of $G[\component_G(u,W)]$.
If this tree were not minimum, replacing it with a lighter spanning tree of $G[\component_G(u,W)]$ would yield a spanning tree of $G$ of strictly smaller weight, again contradicting the minimality of $T$.
\end{proof}

\subsection{Extension cost and touring cost}

To bound the total cost incurred by the algorithm, we distinguish two types of cost.
We call an edge traversed in line~11 and traversed back in line~13 an \emph{extension edge} (blue edges in~\cref{fig:hierarchical-dfs}).
The cost of following the TSP tours computed in line~8 and traversed in line~14 is called \emph{touring cost} (red tours in~\cref{fig:hierarchical-dfs}).
The next two lemmata bound the total cost of traversing extension edges and the total touring cost, respectively.
Note that these traversals account for all non-zero cost incurred by the algorithm.

\begin{lemma}
\label{lem:extension}
The total cost incurred by traversing extension edges is at most $2b\cdot w(T)$.
\end{lemma}

\begin{proof}
For each $W\in b^{\mathbb Z}$, let $E^{(W)}$ be the set of extension edges with weight in $(W/b,W]$, and let $T^{(W)}$ be the set of edges of $T$ in the same weight range.
We prove the stronger claim that $|E^{(W)}|=|T^{(W)}|$.
The lemma then follows by summing over all $W$, since every edge of~$E^{(W)}$ has weight at most $b$ times that of any edge of $T^{(W)}$.
The factor~$2$ accounts for the two traversals in lines~11 and~13.

To prove the claim, fix a component $C$ of $G_W$, and let $k$ be the number of components of~$G_{W/b}$ contained in $C$.
We show that the edges of $E^{(W)}$ and $T^{(W)}$ within $C$ each form a tree on these $k$ components, implying that they both have size $k-1$.
By \cref{lem:components_mst}, $T_W[C]$ is a tree and $T_{W/b}$ has the same $k$ components in $C$ as $G_{W/b}$, so that $T_W$ forms a tree on these components.
For~$E^{(W)}$, recall that every extension edge is a boundary edge when first traversed.
By \cref{obs:algorithm_basics} (\ref{item:obs_fully_explore}, once a component of $G_{W/b}$ is first visited, it is fully explored before an edge of weight greater than~$W/b$ is traversed, i.e., before the agent leaves it again.
Thus every edge of $E^{(W)}$ leads from an already visited component of $G_{W/b}$ to a previously unvisited one.
These edges therefore also form a tree on the $k$ components (they form a depth-first search tree, cf.~\cref{fig:hierarchical-dfs}).

The two edge sets thus contain the same number of edges within $C$.
Summing over all components $C$ of $G_W$ proves the claim.
\end{proof}

Next, we turn to bounding the touring cost.

\begin{lemma}
\label{lem:touring}
For $b>1$, the total touring cost is at most $2(\log_b n+2)\,w(T)$.
\end{lemma}

\begin{proof}
First, observe that, for each fixed $W\in b^\Z$ and each component $C$ of $G_{W/b}$, at most one tour of $C$ is computed by a call with threshold~$W$.
Indeed, when such a tour is computed at a vertex~$u$, the component $C$ is already fully explored by \cref{obs:algorithm_basics}.
Every call in line~12 starts at a newly visited vertex, so no subsequent such call can start within $C$.
Moreover, the call at $u$ fully explores $\component_G(u,W)$, so threshold $W$ cannot subsequently be selected again in line~4 at $u$.

By \cref{lem:components_mst}, $T[C]=T_{W/b}[C]$ is a minimum spanning tree of $G[C]$, so the tour of $C$ costs at most $2w(T_{W/b}[C])$.
Summing over all components of $G_{W/b}$, the touring cost incurred by calls with threshold $W$ is therefore at most $2w(T_{W/b})$.

Next, since $\component_G(u,w(T))=V(G)$, any finite threshold selected in line~4 is less than~$b\,w(T)$.
Therefore, whenever the algorithm is invoked with a finite call threshold $W$, we have $W<b\,w(T)$.
The initial call $\Explore_b(s,\infty)$ additionally performs one final tour of the entire graph, costing at most $2w(T)$.
We obtain that the total cost of touring is at most
\begin{equation*}
\text{Touring cost } \leq 2w(T)+ \sum_{W\in b^\Z, W< w(T)} 2w(T_W).
\end{equation*}
Next, we have
\begin{align*}
\sum_{W\in b^\Z, W< w(T)} w(T_W)
&= \sum_{e\in T} w(e) \cdot |\{W\in b^\Z: w(e)\leq W< w(T)\}| \\
&\leq \sum_{e\in T, w(e)>0} w(e) \left( 1+ \log_b\left( \frac{w(T)}{w(e)}\right)\right).
\end{align*}
To estimate this further, we make use of (a special case of) Jensen's inequality, which states that, if $f\colon I \to \R$ is a concave function for some interval $I\subseteq \R$, then for $x_1,\dots, x_k\in I$, we have~$\sum_{i=1}^{k}f(x_i)\leq k f(\sum_{i=1}^{k}x_i/k)$.
Since the function $f(x)=x\log(w(T)/x)$ is concave and~$T$ contains $n-1$ edges, we obtain the bound
\begin{align*}
\sum_{e\in T, w(e)>0} w(e) \log_b\left( \frac{w(T)}{w(e)}\right)
&\leq (n-1) \frac{\sum_{e\in T}w(e)}{n-1} \log_b \left(\frac{w(T)}{(\sum_{e\in T}w(e))/(n-1)}\right)\\
&= w(T) \log_b(n-1) \leq w(T)\log_b(n)
\end{align*}
Combining these bounds, the total touring cost is at most $2w(T)+2w(T)(1+\log_b n)$, as claimed.
\end{proof}

Note that \Cref{lem:extension,lem:touring} together imply \cref{thm:hDFS_b}, where we used that $\log_b(n)=\log(n)/\log(b)$.

\begin{figure}[t]
\centering
\begin{tikzpicture}[
  font=\small,
  lbedge/.style={draw=black!80,line width=.7pt},
  lbvertex/.style={circle,draw=black,fill=black,inner sep=0pt,minimum size=4pt},
  lbweight/.style={fill=white,inner sep=1pt},
  lbbrace/.style={draw=black!70,line width=.55pt,decorate,decoration={brace,mirror,amplitude=4pt}}
]

\begin{scope}
  \coordinate (s) at (0,0.3);
  \coordinate (a) at (1.7,0.3);
  \coordinate (c) at (1.7,1.55);
  \coordinate (d) at (0,1.55);
  \draw[lbedge] (s) -- node[lbweight,below=3pt] {$b^2-\varepsilon$} (a);
  \draw[lbedge] (a) -- node[lbweight,right=4pt] {$0$} (c);
  \draw[lbedge] (c) -- node[lbweight,above=3pt] {$b+\varepsilon$} (d);
  \draw[lbedge] (d) -- node[lbweight,left=4pt] {$1$} (s);
  \foreach \x in {s,a,c,d}
    \node[lbvertex] at (\x) {};
  \node[below left=3pt] at (s) {$s$};
\end{scope}

\begin{scope}
  \foreach \name/\x in {l0/3.8,l1/4.75,l2/5.35,l3/5.9,l4/6.45,s/6.95,p1/7.5,p2/8.05,u/9.75,r1/10.25,r2/10.8,r3/11.35,r4/11.95,r5/12.9}
    \coordinate (\name) at (\x,.95);

  \draw[lbedge] (l0) -- node[lbweight,above=7pt, anchor=west, xshift=-3pt] {$b^{\lfloor\log_b k\rfloor}$} (l1);
  \node at (5.05,.95) {$\cdots$};
  \draw[lbedge] (l2) -- node[lbweight,above=7pt, anchor=west, xshift=-3pt] {$b^3$} (l3);
  \draw[lbedge] (l3) -- node[lbweight,above=7pt, anchor=west, xshift=-3pt] {$b^2$} (l4);
  \draw[lbedge] (l4) -- node[lbweight,above=7pt, anchor=west, xshift=-3pt] {$b$} (s);

  \draw[lbedge] (s) -- node[lbweight,above=3pt] {$1$} (p1);
  \draw[lbedge] (p1) -- node[lbweight,above=3pt] {$1$} (p2);
  \draw[lbedge] (p2) -- node[lbweight,above=3pt] {$1$} (8.35,.95);
  \node at (8.65,.95) {$\cdots$};
  \draw[lbedge] (8.95,.95) -- node[lbweight,above=3pt] {$1$} (u);

  \draw[lbedge] (u) -- node[lbweight,above=7pt, anchor=west, xshift=-3pt] {$b$} (r1);
  \draw[lbedge] (r1) -- node[lbweight,above=7pt, anchor=west, xshift=-3pt] {$b^2$} (r2);
  \draw[lbedge] (r2) -- node[lbweight,above=7pt, anchor=west, xshift=-3pt] {$b^3$} (r3);
  \node at (11.65,.95) {$\cdots$};
  \draw[lbedge] (r4) -- node[lbweight,above=7pt, anchor=west, xshift=-3pt] {$b^{\lfloor\log_b k\rfloor}$} (r5);

  \foreach \x in {l0,l1,l2,l3,l4,s,p1,p2,u,r1,r2,r3,r4,r5}
    \node[lbvertex] at (\x) {};
  \node[below=1pt] at (s) {$s$};

  \draw[lbbrace] (3.8,.55) -- node[midway,below=6pt] {length $O(k)$} (6.93,.55);
  \draw[lbbrace] (6.97,.55) -- node[midway,below=6pt] {$k$ edges of weight 1} (9.73,.55);
  \draw[lbbrace] (9.77,.55) -- node[midway,below=6pt] {length $O(k)$} (12.9,.55);
\end{scope}
\end{tikzpicture}
\caption{Instances witnessing tightness of the bound $\bigO(b+\log_b(n))$ in \cref{thm:hDFS_b}, for $b\geq2$ and some sufficiently small $\varepsilon>0$.
In both graphs, exploration starts at~$s$.
In the left graph, $\Explore_b$ achieves a competitive ratio of $\Omega(b)$, in the right, a competitive ratio of~$\Omega(\log_b(k))=\Omega(\log_b(n))$.}
\label{fig:lower_bound}
\end{figure}
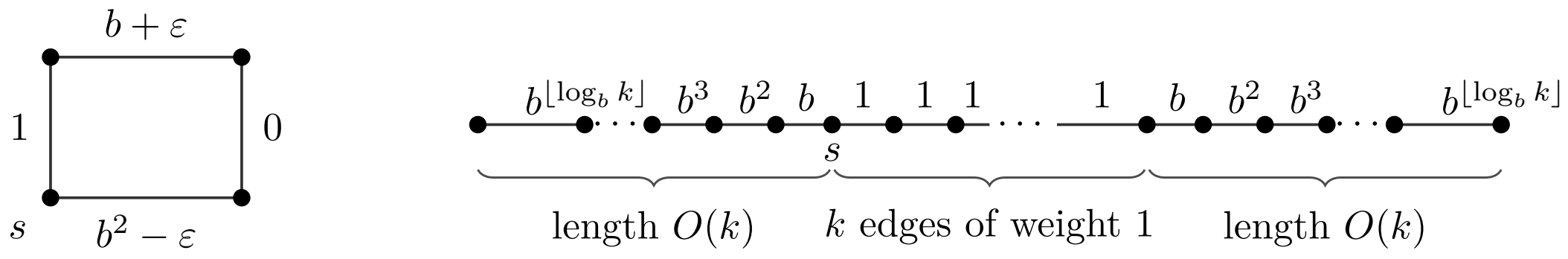

\begin{remark}
The bound in \cref{thm:hDFS_b} is tight.
For sufficiently small $\varepsilon>0$, on the left-hand graph in \cref{fig:lower_bound}, $\Explore_b(s,\infty)$ incurs cost at least $b^2-\varepsilon+1\geq b^2$, while the offline optimum has cost at most $2(1+b+\varepsilon)$.
This yields a competitive ratio of $\Omega(b)$.

On the right-hand graph, which corresponds to the lower bound construction given in~\cite{Megow2012} for hDFS, we have $n=\Theta(k)$, and the offline optimum equals twice the total length of the path, which is $\Theta(k)$.
During the execution of $\Explore_b(s,\infty)$, the agent traverses the~$k$-edge unit-weight subpath at least $\lceil\log_b k\rceil$ times.
Consequently, its competitive ratio is~$\Omega(\log_b k)=\Omega(\log_b n)$.

The two constructions together therefore give a lower bound of $\Omega(b+\log_b n)$ on the competitive ratio of $\Explore_b$, matching the bound in \cref{thm:hDFS_b}.
\end{remark}

\section{Doubling strategy for an unknown number of vertices}
\label{sec:doubling}

While the previous section already shows our main theorem if the agent is given the number of vertices $n$ in advance, we show in this section that a competitive ratio of $\bigO(\log n/ \log \log n)$ can also be achieved without this assumption.

To this end, consider \cref{alg:unknown-n}. 
The algorithm makes a ``guess'' $N$ for the total number of vertices in the graph.
It then sets $b=\sqrt{\log N}$ (if $N$ is large enough), which is the desired choice for~$b$ if the number of vertices is indeed $N$ and executes $\Explore_b$.
When more than~$N$ vertices are learned, it interrupts the algorithm and updates its guess to~$N^2$.
Here, a vertex is called \emph{learned} when it is either explored or a boundary vertex. After interrupting and updating the parameters, the exploration begins again from scratch, i.e., we let the agent ``forget'' about the vertices learned in the previous phase.

\newcommand{\algunknownn}{\ensuremath{\textsc{Adaptive-hDFS}}\xspace}

\begin{algorithm}[h]
\caption{\algunknownn}
\label{alg:unknown-n}
\small
%


$N\gets2$\\
\While{$G$ is not fully explored}{
    $b\gets\max\{2,\sqrt{\log N}\}$\\
    run $\Explore_b(s,\infty)$ and interrupt when more than $N$ vertices are learned\\
	return to $s$ via a shortest known path\\
    $N\gets N^2$\\
}
\end{algorithm}

\begin{figure}[t]
\centering
\begin{tikzpicture}[
  scale=.9,
  font=\small,
  gnknown/.style={draw=black!75,line width=.65pt},
  gnadded/.style={draw=gnblue,line width=1.1pt},
  gnunseen/.style={draw=black!30,line width=.65pt,densely dotted},
  gnexplored/.style={circle,draw=black,fill=black,inner sep=0pt,minimum size=4pt},
  gnboundary/.style={circle,draw=black,fill=white,line width=.8pt,inner sep=0pt,minimum size=5.5pt},
  gnhidden/.style={circle,draw=black!25,fill=black!25,inner sep=0pt,minimum size=4pt},
  gnzero/.style={text=gnblue,fill=white,inner sep=1pt,font=\footnotesize}
]
\definecolor{gnblue}{RGB}{0,114,178}

\def\gnbase{%
  \coordinate (s) at (-.9,-.4);
  \coordinate (a) at (-.55,.65);
  \coordinate (b) at (.45,.7);
  \coordinate (c) at (.95,-.2);
  \coordinate (d) at (0,-.75);
  \coordinate (v) at (1.874,-.112);
  \coordinate (p) at (1.506,1.492);
  \coordinate (q) at (-.638,1.618);
  \coordinate (r) at (-1.824,-1.236);
  \coordinate (t) at (.792,-1.762);
  \foreach \x/\y in {s/a,a/b,b/c,c/d,d/s,a/d,b/p,a/q,s/r,d/t,c/v}
    \draw[gnknown] (\x) -- (\y);
}

\def\gnvertices{%
  \foreach \x in {s,a,b,c,d}
    \node[gnexplored] at (\x) {};
  \foreach \x in {p,q,r,t}
    \node[gnboundary] at (\x) {};
  \node[gnboundary,draw=gnblue,line width=1.2pt,minimum size=6.5pt] at (v) {};
  \node[anchor=west,inner sep=5pt] at (v) {$v$};
  \node[anchor=south east,inner sep=3pt] at (s) {$s$};
}

\begin{scope}
  \gnbase
  \coordinate (h1) at (-1.35,2.45);
  \coordinate (h2) at (-2.85,-1.95);
  \coordinate (h3) at (1.65,-2.6);
  \coordinate (h4) at (2.85,-1.35);
  \coordinate (h5) at (2.75,2.35);
  \foreach \x/\y in {q/h1,r/h2,t/h3,v/h4,p/h5,h1/h5,h3/h4}
    \draw[gnunseen] (\x) -- (\y);
  \draw[gnunseen] (h2) .. controls (-1.6,-2.85) and (.5,-3.0) .. (h3);
  \foreach \x in {h1,h2,h3,h4,h5}
    \node[gnhidden] at (\x) {};
  \gnvertices
\end{scope}

\begin{scope}[xshift=7cm]
  \gnbase
  \draw[gnadded] (v) -- node[gnzero,midway,right=2pt] {$0$} (p);
  \draw[gnadded] (v) -- node[gnzero,pos=.55,right=2pt] {$0$} (t);
  \draw[gnadded] (v) .. controls (3.35,2.3) and (.3,3.0) .. node[gnzero,pos=.58,above=2pt] {$0$} (q);
  \draw[gnadded] (v) .. controls (3.05,-2.65) and (-.6,-3.0) .. node[gnzero,pos=.6,below=2pt] {$0$} (r);
  \gnvertices
\end{scope}


\end{tikzpicture}
\caption{Construction of the auxiliary graph $G_N$ from the state immediately before visiting~$v$.
The left panel shows the exploration state immediately before visiting $v$, whose visit causes the number of learned vertices to exceed $N$. Here, filled vertices are explored, hollow vertices are boundary vertices, gray vertices and dotted edges are unrevealed.
On the right, $G_N$ is depicted. The blue edges are edges not (necessarily) present in $G$.}
\label{fig:auxiliary-graph}
\end{figure}
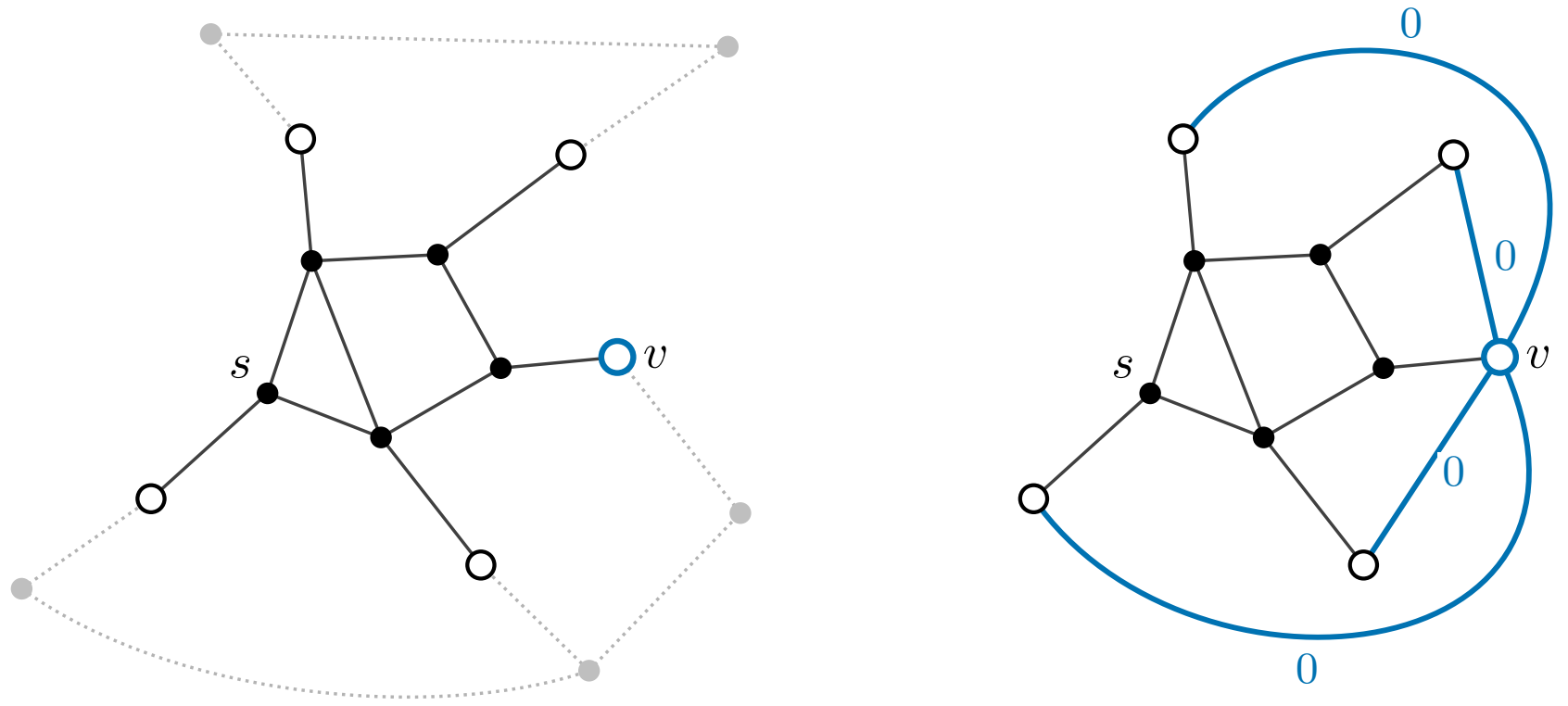

Explained on an intuitive level, our main result follows because each update of $N$ increases the phase competitive ratio of $\log N/ \log \log N$ by a constant factor (for $N$ large enough), so the sum of these bounds is dominated by the last one.
We work out the details in the following theorem.

\begin{theorem}
\label{thm:unknown-n}
The algorithm $\algunknownn$ is $\bigO(\log n/\log\log n)$-competitive.
\end{theorem}

\begin{proof}
Consider exploration of a graph $G$ and let $T$ be a minimum spanning tree of $G$. In the following, we call one iteration of the while-loop a \emph{phase}.
Consider an interrupted phase of positive cost with guess $N$ and parameter $b=\max\{2,\sqrt{\log N}\}$, and let $v$ be the vertex whose visit causes more than $N$ vertices to be learned.
We construct a graph $G_N$ as follows (cf.~\cref{fig:auxiliary-graph}): it consists of all vertices and edges learned immediately before visiting $v$ (in particular, $v$ was a boundary vertex at that time), and we add edges of weight~0 from $v$ to all other boundary vertices.
Then~$G_N$ has at most $N$ vertices and, crucially,  $\Explore_b$ has the same exploration trajectory on~$G_N$ up to the arrival at $v$ as on $G$.
Therefore, by \cref{thm:hDFS_b}, the cost before interruption is at most~$\bigO(b+\log N/\log b)\opt(G_N)$.
Returning to $s$ at most doubles this cost.

The learned edges of $T$, together with the added zero-weight edges, form a connected spanning subgraph of $G_N$ of weight at most $w(T)$:
Indeed, every learned vertex is connected to $v$ by following its unique path in $T$ to the first boundary vertex and then using a zero-weight edge to $v$.
Thus,~$\opt(G_N)\leq 2 w(T)$ and the total cost incurred in a phase with parameters~$N$ and $b$ is at most~$\bigO(b+\log N/\log b) w(T)$.
 
We can thus bound all cost incurred in the first two phases by a constant times $\opt(G)$, which we hide in the $\bigO$-notation for the desired competitive ratio.
From the third phase on, we have $N\geq 16$ so that $b=\sqrt{\log N}$, and the cost of such a phase is bounded by~$\bigO(\log N/\log\log N)w(T)$.

The guess in phase $i$ is $N=2^{2^i}$ and $\log N/\log \log N=2^i/i$. The algorithm terminates once the guess is at least $n$, i.e., in phase $i$ for $i=\lceil \log \log n \rceil$.
Absorbing the first two phases into the constant term hidden in the $\bigO$-notation, we obtain the bound
\begin{align*}
\algunknownn(G,s)
&\leq
\bigO\left(1+\sum_{i=2}^{\lceil\log \log n\rceil}\frac{2^i}{i}\right)w(T)
=\bigO\left(\frac{\log n}{\log\log n}\right)w(T).
\end{align*}
Here, the sum is bounded by a constant multiple of its last term.
Since $w(T)\leq\opt(G)$, this proves the claimed competitive ratio.
\end{proof}

\paragraph*{AI disclosure.}
The result presented in this paper was developed with the assistance of ChatGPT~6 Astra through discussions of ideas for improved algorithms for online graph exploration.
Following these discussions, the author independently verified the mathematical arguments, simplified the proof, and prepared the exposition.
The author assumes full responsibility for the mathematical correctness of all presented arguments and the quality of the exposition.


\bibliographystyle{alphaurl}
\bibliography{hDFSSublogarithmic}

\end{document}